\documentclass[seceqno,cite]{cimento}

\usepackage{pstricks}
\usepackage{pst-node}
\usepackage{pst-fill}
\usepackage{pst-coil}
\usepackage{yhmath}
\usepackage{mathrsfs}
\usepackage{amssymb}
\usepackage{amsmath}
\usepackage{graphicx}

\def\={\;\;=\;\;}                               
\def\df{\frac{}{}}                              

\renewcommand{\mathbf}[1]{\boldsymbol{#1}}

\def\w{\wedge}                                  
\def\wt#1{\widetilde{#1}}                       
\def\Lie{\mathcal{L}}                           
\def\pdiff#1#2{\frac{\partial#1}{\partial#2}}   
\def\({\left(}                                  
\def\){\right)}                                 
\def\[{\left[}                                  
\def\]{\right]}                                 
\def\RR{{\mathbb R}}                            
\def\sect{{\cal{S}}}                            
\def\conj#1{\overline{#1}}                      

\def\Re{\text{Re}}                              

\def\ep#1{\epsilon_{#1}}                        
\def\qquadand{\qquad\text{and}\qquad}

\def\EU{{\cal E \, }}

\def\Me{{\mathbf e}}
\def\Mb{{\mathbf b}}
\def\Md{{\mathbf d}}
\def\Mh{{\mathbf h}}
\def\bfE{\mathbf{E}}
\def\bfB{\mathbf{B}}
\def\bfD{\mathbf{D}}
\def\bfH{\mathbf{H}}
\def\bbfj{{\bf j}}
\def\bbfJ{{\bf J}}
\def\D{\widehat{d}}
\def\cc{\, c_0 \,}

\def\PermDE{{\zeta^{\textup{de}}}}
\def\PermDB{{\zeta^{\textup{db}}}}
\def\PermHE{{\zeta^{\textup{he}}}}
\def\PermHB{{\zeta^{\textup{hb}}}}
\def\GenPoyn{{s}}
\def\TT{{\cal T}}

\def\Sig{{\Sigma^3}_t}
\def\Sigs{{\Sigma^3}_s}

\def\JK{{J}^{U}_K}
\def\RK{{\rho}^{U}_K}

\def\rr{{\boldsymbol r}}

\def\kk{{\boldsymbol k}}
\def\KK{{{\cal K}_K}}
\def\KKK{{\boldsymbol K}}

\def\SKK{{\sigma_K}^{U\,\,{field}\,\, I_j}}
\def\SKKK{{\sigma_K}^{U\,\,rem\, }}

\def\TK{{{\tau}_K}}
\def\TKK{{\TK}^{{field}\,\, I_j}}
\def\TKKK{{\TK}^{rem\,\, }}

\def\TEM{{\tau_K}^{EM}}
\def\SEM{{\sigma_K}^{U\,\,EM}}

\def\Vj{{V}^{I_j}}

\def\Id{I\!d}

\def\Y{{\kk, \omega}}

\def\FA{\check{\alpha}_{\Y}}

\def\Fnn{\check{\boldsymbol {n}}_{\Y}}
\def\Fnnr{\check{\boldsymbol {n}}_{\Y}^{r}}

\def\Fee#1#2{\check{\boldsymbol e}_{\Y}^{#1 \, #2}}   
\def\Fbb#1#2{\check{\boldsymbol b}_{\Y}^{#1 \, #2}}
\def\Fdd#1#2{\check{\boldsymbol d}_{\Y}^{#1 \, #2}}
\def\Fhh#1#2{\check{\boldsymbol h}_{\Y}^{#1 \, #2}}

\def\FFee{\check{e}_{\Y}}

\def\FBB#1#2{\check{\boldsymbol B}_{\Y}^{#1 \, #2}}   
\def\FDD#1#2{\check{\boldsymbol D}_{\Y}^{#1 \, #2}}

\def\Xde{\check{\boldsymbol \zeta}^{de}_{\Y}}
\def\Xdb{\check{\boldsymbol \zeta}^{db}_{\Y}}
\def\Xhe{\check{\boldsymbol \zeta}^{he}_{\Y}}
\def\Xhb{\check{\boldsymbol \zeta}^{hb}_{\Y}}

\def\XD{\check{ {\cal D}}_{\Y}}
\def\XDD{\check{\boldsymbol {\cal D}}_{\Y}}

\def\Fk#1#2#3{{k^{#1 #3}_#2}  }        
\def\FE#1#2#3{{E^{#1\,#3}_#2 }  }      

\title{An Intrinsic Approach to Forces in Magnetoelectric Media}

\author{R. W. Tucker\from{ins:a}\thanks{email address:\;r.tucker@lancaster.ac.uk} \atque T.J.Walton\from{ins:a}\thanks{email address:\;t.walton2@lancaster.ac.uk}}

\instlist{\inst{ins:a} Department of Physics, Lancaster University and The Cockcroft
Institute}

\PACSes{ \PACSit{02.40.Hw}{Classical differential geometry}
         \PACSit{03.50.De}{Classical electromagnetism}
         \PACSit{41.20.-q}{Applied classical electromagnetism}
         \PACSit{41.20.Jb}{Electromagnetic wave propagation}
         \PACSit{46.05.+b}{General theory of continuum mechanics of solids}
        }

\begin{document}
\maketitle

\begin{abstract}
    This paper offers a conceptually straightforward method for the
    calculation of stresses in polarisable media based on the notion
    of a {\it drive form} and its property of being closed in
    spacetimes with symmetry. After an outline of the notation
    required to exploit the powerful exterior calculus of differential
    forms, a discussion of the relation between Killing isometries and
    conservation laws for smooth and distributional drive forms is
    given. Instantaneous forces on isolated spacetime domains
    and regions with interfaces are defined, based on manifestly
    covariant equations of motion. The remaining sections apply these
    notions to media that sustain electromagnetic stresses, with
    emphasis on homogeneous magnetoelectric material. An explicit
    calculation of the average pressure exerted by a monochromatic
    wave normally incident on a homogeneous, magnetoelectric slab in vacuo
    is presented and the concluding section summarizes how this pressure depends
    on the parameters in the magnetoelectric tensors for the medium.
\end{abstract}

\section{Introduction}
\label{ch1}
The calculation of stresses in material media has extensive
application in modern science. The balance laws of continuum
mechanics offer an established framework for such calculations for
matter subject to a wide class of constitutive properties that
attempt to accommodate interaction with the environment in terms
of phenomenological relations \cite{Maugin}, \cite{Landau}. Such
relations are not always easily accessible via experiment, since
the response of matter to internal and external interactions can
be very complex. If one formulates these interactions in the
language of forces derived from stress-energy-momentum tensors,
then it is sometimes non-trivial to determine experimentally an
appropriate tensor that can be associated with a particular class
of interactions on a macroscopic scale
\cite{Barnett}$-\!$\cite{Dietz}. This problem has led to numerous
debates over the last century about how best to formulate the
transmission of electromagnetic forces in polarisable media. Since
the electromagnetic interaction is fundamentally relativistic in
nature, the problem is compounded if one insists on a
relativistically (covariant) theoretical formulation to compare
with experiment in the laboratory. Judged by the large literature
on this subject, there is no universal consensus on how best to
calculate forces in polarisable media and hence the needed
experimental input into the subject has been of uncertain value in
the past. However, modern technology $-$ with the refined
experimental procedures now available $-$ offers the possibility
that the appropriate constitutive relations for certain classes of polarisable matter
can be determined experimentally \cite{Hinds} over a broad
range of field intensities, frequencies and geometric configurations. Furthermore, new materials with novel
constitutive properties are  being fabricated \cite{Sim} and
their response to time-varying electromagnetic fields also offers
new potential for technological advances. With these points in
mind, this paper offers a conceptually straightforward method for
the calculation of stresses in polarisable media, based on the
notion of a {\it drive form} and its property of being closed in
spacetimes with symmetry. Section 2 outlines the notation required
to exploit the powerful exterior calculus of differential forms
that is used throughout the article. Sections 3 and 4 relate the
isometries to conservation laws for smooth and distributional
drive forms. Sections 5 and 6  discuss equations of motion in spacetime
and how they may be used to define instantaneous forces on
isolated domains, while section 7 deals with forces on domains
with interfaces. The remaining sections apply these notions to
media that sustain electromagnetic stresses, with emphasis in
section 11 on homogeneous, magnetoelectric material. In sections
12$-\!$14, an explicit calculation of the average pressure exerted
by a monochromatic, electromagnetic wave on a homogeneous,
magnetoelectric slab in vacuo is presented and the discussion in
section 15 summarises how this pressure depends on the parameters
of the magnetoelectric tensors for the medium.\\

\section{Notation}
\label{ch2}
The formulation below exploits the geometric language of exterior
differential forms and vector fields on a manifold $M$ \cite{RWT}.
Such a language is ideally suited to accommodate {\it local}
changes of coordinates that can be used to simplify the
description of boundary value problems and naturally encapsulates
intrinsic {\it global} properties of domains with different
physical properties. It is also makes available the powerful
exterior calculus that facilitates the integration of forms over
domains described as the images of chain maps and permits a clear
formulation of notions such as energy, momentum, angular momentum,
force and torque, by fundamentally relating them to isometries of
spacetime. In this framework, a $p$-form $\alpha$ belongs to
$\sect\Lambda^p M$, the space of sections of the bundle of
exterior $p$-forms over $M$, while vector fields $X$ belong to
$\sect TM$, the space of sections of the tangent bundle over $M$.
On a manifold with metric tensor $g$, we denote $g(X,-)$ by
$\wt{X}\in \sect\Lambda^1 M$ and conversely set $\wt{\wt{X} }=X$
for all $X$. In the following, a notational distinction between
smooth $(C^\infty)$ forms on some regular domain and those with
possible singularities or discontinuities is useful. Smooth forms
with compact support on spacetime will be referred to as {\it test
forms} \cite{Shilov} and distinguished below by a superposed hat.
Manifolds with dimension $n$ will be assumed orientable and
endowed with a preferred $n$-form induced from the metric tensor
field $g$. One then has \cite{RWT} the linear Hodge operator
$\star$ that maps $p$-forms to $(n-p)$-forms on $M$. If $g$ has
signature $t_g$, one may write
\begin{eqnarray}
    g=\sum_{i=1}^n e^i \otimes e^j \,\eta_{ij},
\end{eqnarray}
where $\eta_{ij}=diag(\pm 1,\pm 1,\ldots \pm 1)$ and
\begin{eqnarray}
    \star 1= e^1 \w e^2 \w \ldots \w e^n,
\end{eqnarray}
with $t_g=det(\eta_{ij})$ and $\{ e^i\}$ a set of basis 1-forms in
$\sect\Lambda^1M$. The natural dual basis $\{ {X_i}\}$ is defined
so that $ e^i ( {X_j} )=\delta^i_j$ and the contraction operator
with respect to $X$ is denoted $i_{X}$. Covariant differentiation
is performed with respect to the metric compatible Levi-Civita
connection
$\nabla$, whilst Lie differentiation is denoted $\Lie$.\\

\section{Isometries and Drive Forms}
\label{drive}
The notion of a {\it drive form} arises from the theory of
gravitation in spacetimes $M$ with isometries. In Einstein's
theory of gravitation, the metric $g$ of spacetime is determined
by the tensor field equation
\begin{eqnarray*}
    Ein=\TT,
\end{eqnarray*}
where $Ein\in\sect T^2 M$ denotes the degree 2 symmetric
divergence-free Einstein tensor field. Hence $\TT$ must be a
symmetric divergence-free degree 2 tensor field:
\begin{eqnarray*}
    \nabla\cdot\TT=0.
\end{eqnarray*}
The tensor $\TT$ is regarded as a source of gravitational
curvature\footnote{The tensor $\TT$ has dimensions of $[MLT^{-2}]$
(force) constructed from the SI dimensions $[M], [L], [T], [Q]$
where $[Q]$  has the unit of the Coulomb in the MKS system.}. If
$K$ is a Killing vector field generating a spacetime symmetry and
$\star$ is the Lorentzian Hodge operator associated with $g$, then
by definition
\begin{eqnarray*}
    \Lie_K g=0
\end{eqnarray*}
and it follows that the {\it drive 3-form}
\begin{eqnarray*}
    \tau_K &\equiv& \star (\TT(K,-))
\end{eqnarray*}
is closed on some domain $I_j$ of $M$:
\begin{eqnarray*}
    d\,\tau_K=0.
\end{eqnarray*}
If the spacetime admits a set of Killing vector fields $\{K_i \in
\sect T\,I_j\}$, one has a conservation law for each $K_i$ in
every regular spacetime domain $I_j$ \cite{RWT}, \cite{Benn}.
These may be supplemented with (tensor or spinor) field equations
\begin{eqnarray*}
    {\cal E}^{I_j}(g,\Phi_\alpha^{I_j})=0,
\end{eqnarray*}
for all piecewise smooth (tensor or spinor) fields
$\Phi_\alpha^{I_j}$ that interact with each other and gravity.
These field equations may induce compatibility conditions and
further (non-Killing) conservation laws
\begin{eqnarray*}
    d\,{\cal J}^{I_j}(\Phi_\alpha^{I_j})=0
\end{eqnarray*}
(e.g. electric charge-current conservation). In phenomenological
models, some of the field equations may be replaced by fixed
background fields and source currents, together with consistent
constitutive relations between these fields and currents.

An observer field is associated with an arbitrary {\it unit}
future-pointing timelike 4-velocity vector field $U\in\sect TM$.
The field $U$ may be used to describe an {\it observer frame} on
spacetime and its integral curves model {\it idealized observers}.
The {drive form} $\tau_K$ associated with any $K$ admits a unique
orthogonal decomposition with respect to any observer frame $U$:
\begin{eqnarray*}
    \tau_K=\JK\w \wt{U} + \RK,
\end{eqnarray*}
where the spatial forms $\RK\in \sect\Lambda^3 M$ and $\JK\in
\sect\Lambda^2 M$ satisfy $i_U\RK=i_U\JK=0$. In a local region,
the conservation law $d\,\tau_K=0$ implies, in terms of the
$K\!$-current $\JK$, the continuity relation in the frame $U$:
\begin{eqnarray*}
    d\,\JK+\Lie_U\tau_K=0.
\end{eqnarray*}
If $K$ is a {\it spacelike translational} Killing vector field and
$U$ a unit time-like (future-pointing) 4-vector observer field
\footnote{The frame is inertial if $\nabla U=0$.}, then
\begin{eqnarray*}
    \JK\equiv -\, i_U\tau_K
\end{eqnarray*}
is the linear momentum current (stress) 2-form in the frame $U$
and
\begin{eqnarray*}
    \RK\equiv -(i_U\star\tau_K)\star \wt{U}
\end{eqnarray*}
is the associated linear momentum density 3-form in the frame $U$.
If $K$ is a {\it spacelike rotational} Killing vector field
generating $SO(3)$ group isometries, then $\JK$ is an
angular-momentum current (torque stress) 2-form and $\RK$ is the
associated angular-momentum density 3-form in the frame $U$. If
$K$ is a {\it timelike translational} Killing vector field, then
$\JK$ is an energy current (power) 2-form and $\RK$ is the
associated energy density 3-form in the frame $U$. In the
following, attention is restricted to translational spacelike
Killing vectors of flat spacetime and the computation of integrals
of $\JK$ for a particular contribution to $\tau_K$ associated with
electromagnetic fields in homogeneous but anisotropic media of a
particular kind. It will be argued that this formulation leads to
a natural definition of integrated static forces in media with
discontinuous material behavior and highlights the need for care
in giving a practical definition of integrated force in media in
the presence of time varying fields.\\

\section{Distributional Drive Forms}
\label{drive}
To accommodate media with singular time-dependent sources of
stress (e.g at surface interfaces or lines in space), introduce
the distributional Killing 3-form $\TK^D$ on spacetime and its
distributional source $\KK^D$ satisfying the distributional
equation
\begin{eqnarray}\label{distrib}
    d\,\TK^D[\hat\beta]=\KK^D[\hat\beta],
\end{eqnarray}
for all test 4-forms $\hat\beta$ \cite{Shilov} on spacetime.
Consider a compact medium at time $t$, with spatial volume
determined by the image of the spacelike $t$-parameterised
immersion $\Sig: W_3\subset \RR^3 \to M$, evolving for a finite
interval of time. Denote its immersed history in spacetime by the
region $I_{1}$. Let $I_{2}$ be a compact region of spacetime
outside this medium history. It follows from (\ref{distrib}) that
if ${\TK}^{I_{1}}$ is the regular drive form in region $I_{1}$ and
${\TK}^{I_{2}}$ is the regular drive form in region $I_{2}$, then
\begin{eqnarray}
    \label{t1} d\,{\TK}^{I_{1}} &=& 0\qquad  \text{in}\quad I_{1}, \\
    \label{t2} d\,{\TK}^{I_{2}} &=& 0\qquad  \text{in}\quad I_{2} \\ \mbox{\,\,and\,\,}\qquad\qquad
    \label{t3} {\Sigs}^{\star}(\TK^{I_{1}} -\TK^{I_{2}} + \KK) &=& 0,
\end{eqnarray}
at an evolving interface defined by the timelike,
$t$-parameterised immersion $\Sigs: S_2\subset \RR^3\to M$ between
$I_{1}$ and $I_{2}$ with a smooth interface drive form $\KK$ on
its image. The history of these images in spacetime is indicated
schematically in figure (1).
\begin{figure}[h]
\setlength{\unitlength}{2.5cm}
\begin{center}
\begin{picture}(6,3.6)
    \qbezier(1,3.6)(0.1,3.6)(0.1,3)
    \qbezier(1,3.6)(1.2,3.65)(2,3.4)
    \qbezier(2,3.4)(2.5,3.22)(3,3.2)
    \qbezier(3,3.2)(3.5,3.24)(4,3.25)
    \qbezier(4,3.25)(5.1,3.2)(5.5,2)
    \qbezier(5.5,2)(5.54,1.9)(5.5,1.5)
    \qbezier(5.5,1.5)(5,0.2)(2.1,0.2)
    \qbezier(2.1,0.2)(0.8,0)(0.4,0.9)
    \qbezier(0.4,0.9)(0,1.9)(0.1,3)

    \psellipse[](6,7)(1.8,.4)
    \pccoil[coilarm=0cm,coilaspect=0,coilheight=6,coilwidth=1.1](4.24,2)(4.24,7)
    \pccoil[coilarm=0cm,coilaspect=0,coilheight=6,coilwidth=1.1](7.8,2)(7.8,7)
    \psframe*[linecolor=white](4.95,1.7)(5.7,2.35)
    \psframe*[linecolor=white](7.3,1.9)(7.8,2.1)
    \put(2,0.75){\mbox{\Large $\Sigma^{3}_{\;\;t_{0}}$}}
    \psframe*[linecolor=white](3.7,1.9)(4.3,2.1)
    \psellipse[fillstyle=vlines,hatchwidth=0.2pt](5.5,2)(1.8,.4)

    \psline[linewidth=1.3pt,arrowsize=3pt 2]{->}(6.8,2)(6.7,2.9)
    \psline[linewidth=1.3pt,arrowsize=3pt 2]{->}(6.05,2)(6,2.9)
    \psline[linewidth=1.3pt,arrowsize=3pt 2]{->}(4.7,2)(4.8,2.9)
    \psline[linewidth=1.3pt,arrowsize=3pt 2]{->}(4,2)(4.2,2.9)
    \put(2.5,1.2){\mbox{\Large $V$}}

    \psframe*[linecolor=white](5.45,6.74)(6.25,7.33)
    \put(2.1,2.75){\mbox{\Large $\Sig$}}
    \psellipse[fillstyle=vlines,hatchwidth=0.2pt](6,7)(1.8,.4)
    \psline[linewidth=1.3pt,arrowsize=3pt 2]{->}(5.9,6.9)(6.7,7.2)
    \psframe*[linecolor=white](6.72,6.84)(7.15,7.23)
    \put(2.7,2.75){\mbox{\large $N$}}

    \psellipse[linestyle=dashed,fillstyle=vlines,hatchwidth=0.2pt](6.45,4.6)(1.8,.4)
    \psellipticarc[](6.45,4.6)(1.8,.4){180}{0}

    \multiput(4,1.3)(0,0.3){4}{\psline[linewidth=1.3pt,arrowsize=3pt 3]{->}(0,0)(1,0)}
    \multiput(4.25,1.3)(0,0.3){4}{\psline[linewidth=1.3pt](0,0)(1,0)}

    \put(4.2,2.4){\mbox{\Large $K$}}

    \put(2.4,2.2){\mbox{\huge $I_{1}$}}

    \psline[linewidth=1pt,arrowsize=3pt 4,linearc=1.2]{->}(2,2.5)(2.8,4)(2.1,5)(2.2,5.6)
    \psline[linewidth=1pt,linearc=1.3](2.2,5.6)(2.4,6.6)

    \psline[linewidth=1pt,arrowsize=3pt 4,linearc=1.2]{->}(1.5,2.5)(2.3,4)(1.6,5)(1.7,5.6)
    \psline[linewidth=1pt,linearc=1.3](1.7,5.6)(1.9,6.6)

    \psline[linewidth=1pt,arrowsize=3pt 4,linearc=1.2]{->}(2.5,2.5)(3.3,4)(2.6,5)(2.7,5.6)
    \psline[linewidth=1pt,linearc=1.3](2.7,5.6)(2.9,6.6)

    \put(1.4,2.2){\mbox{\huge $I_{2}$}}

    \put(0.9,2.75){\mbox{\Large $U$}}

    \put(3.6,0.78){\mbox{\Large $\Sigs$}}

    \psline[linestyle=dashed,dash=2pt 2pt,arrowsize=3pt 4]{->}(9,2.2)(7.5,3)
\end{picture}
\caption{The partition of spacetime $M$ by the history of a
compact medium (with boundary $\Sigs \cup \Sigma^{3}_{\;\;t_{0}}
\cup \Sig$), evolving with 4-velocity $V$. The timelike vector
field $U$ defines a frame, $N$ is a unit, spacelike vector field
and $K$ is a Killing vector field.}
\end{center}
\label{History}
\end{figure}
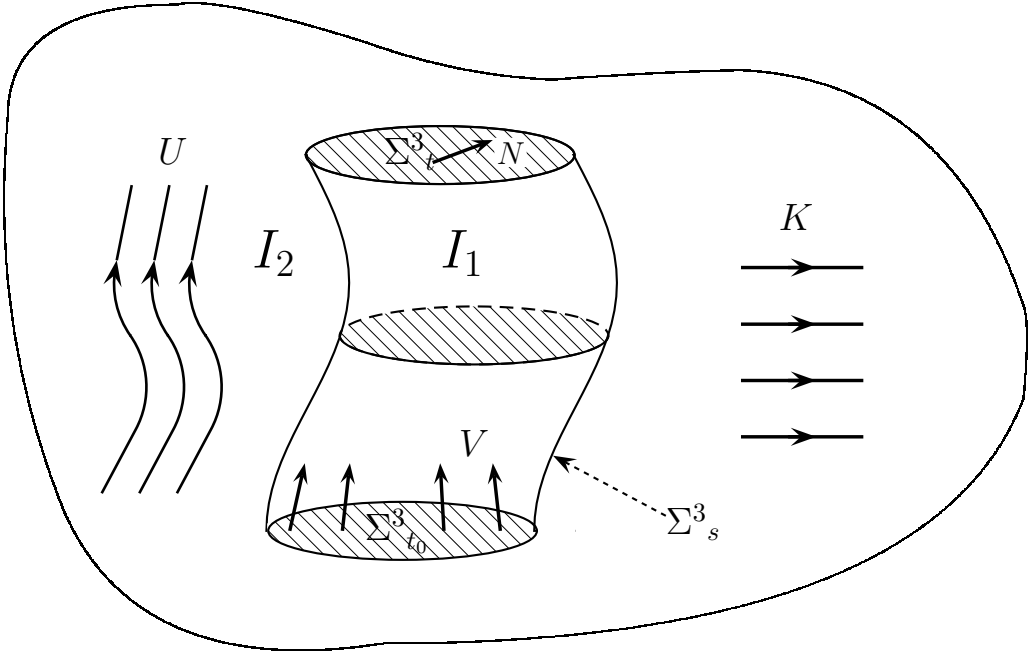

\section{Equation of Motion for a Smooth Domain}
\label{drive}
The notion of force (and torque) is implicit in the balance laws
of classical Newtonian continuum mechanics. In the presence of
time-varying fields, it is natural to associate energy, momentum
and angular momentum with such fields in order to maintain the
conservation of these quantities for closed systems. The only
sensible approach to defining force (and torque) density in such
circumstances, where the balance law arises from the divergence of
a total drive form for the system, is with respect to a particular
splitting of this divergence. For systems without mechanical
constraint, one assigns a smooth 4-velocity $V$ (and angular
velocity) field to each smooth domain to describe the motion. The
jumps in these fields at interfaces between domains must be
computed from (\ref{t1}$-$\ref{t3}) above. The 4-acceleration
field $A$ of each domain (and possibly its rate of change) will
appear in one or more components of the split and the remaining
terms in the divergence are often identified with total force (or
torque) densities for the domain. However, unless one prescribes
how to practically identify component contributions to the total
force (for example by cancelling some of them by externally
applied mechanical constraints), there is no natural way to
identify a canonical split of the divergence of the total drive
form. In those situations where the interaction of matter and
fields is stationary or static, one can appeal to static
experiments with non-moving media to try and give an unambiguous
definition to material body forces. For electromagnetic
interactions with polarisable media, comparison with experiment is
difficult, since the choice of drive form is very model dependent
for many materials. However modern technology $-$ with the refined
experimental procedures now available $-$ offers the possibility
that the appropriate constitutive relation for certain types of  matter
can be determined experimentally over an extended parameter range.
\cite{Sim}.

To illustrate these general remarks, consider an uncharged
(unbounded) medium containing a fixed number of constituents, with
number density ${\cal N} \in \sect\Lambda^0 M$ and mass density
$\rho=m_0{\cal N}$, $m_0>0$, in Minkowski spacetime with mass
conservation $d\,(\rho\star \wt{V})=0$. Write
\begin{eqnarray*}
    \TK={\TK}^V+ {\TK}^{{field}},
\end{eqnarray*}
where $V$ is the (future-pointing) unit, time-like 4-velocity
field of the medium and $K$ a Killing vector field. For a simple
medium with a smooth mass density, suppose
\begin{eqnarray*}
    {\TK}^V\equiv \cc^2\,\rho g(V,K) \star \wt{V},
\end{eqnarray*}
with $\cc$ the speed of light in vacuo, then
\begin{eqnarray*}
    d\,\TK=0
\end{eqnarray*}
yields the local equation of motion \cite{RWT} for the field $V$:
\begin{eqnarray*}
    \cc^2\,\rho\wt{A}(K)=f_K,
\end{eqnarray*}
where $f_K\equiv\star d\,{\TK}^{{field}}$ and the 4-acceleration
1-form $\wt{A}\equiv \nabla_V \wt{V}$. If $\nabla V=0$, then
$\wt{A}=0$ and the motion of the medium is geodesic. The medium is
then static in the frame where $U=V$. Contracting the local
equation of motion
\begin{eqnarray*}
    \cc^2\,\rho \wt{A}(K) \star 1 + d\,{\TK}^{{field}}=0
\end{eqnarray*}
with the observer field $U$ and integrating over
the volume $\Sig$ yields
\begin{eqnarray*}
    \dot{P}_K^{mech\,U}[\Sig]=f_K^U[\Sig],
\end{eqnarray*}
where
\begin{eqnarray*}
    \dot{P}_K^{mech\,U}[\Sig]\equiv{\int}_{\Sig} \mu_U\wt{A}(K).
\end{eqnarray*}
Here the mass 3-form
\begin{eqnarray*}
    \mu_U\equiv -\cc^2\rho\star\wt{U}
\end{eqnarray*}
and the total instantaneous {\it integrated} $K\!$-drive component on
$\Sig$ at time $t$ in the $U$ frame is
\begin{eqnarray*}
    f_K^U[\Sig]\equiv  {\int}_{\Sig} i_U\,d\,{\TK}^{{field}}.
\end{eqnarray*}\\

\section{The General Integrated Force Form on a Regular Domain $I_j\subset M$}
\label{regularforce}
In Minkowski spacetime, one has a global basis of parallel unit
space-like translational Killing vector fields $(K_1,K_2,K_3)$. In
local Cartesian coordinates $\{ t,x,y,z \}$, with ${g=-\cc^2 dt
\otimes dt + dx \otimes dx  +   dy \otimes dy  +  dz \otimes dz}
$:
\begin{eqnarray*}
    K_1=\pdiff{}{x},\,\,K_2=\pdiff{}{y}, \,\,K_3=\pdiff{}{z}.
\end{eqnarray*}
One can then define the instantaneous {\it integrated} force 1-form on
$\Sig$ at time $t$ in the $U$ frame to be
\begin{eqnarray}\label{force1}
    f^U[\Sig]\equiv\sum_{j=1}^3 f_{K_j}^U[\Sig] \,\,\wt{K_j}.
\end{eqnarray}
Then, if $N$ is any unit space-like vector field on $\Sig$, the
instantaneous integrated force component in the direction $N$
acting on $\Sig$ is
\begin{eqnarray}\label{force2}
    f^U[\Sig](N) = \sum_{j=1}^3 f_{K_j}^U[\Sig]\,\,\wt{K_j}(N).
\end{eqnarray}
In an arbitrary (possibly non-inertial) frame $U$ and domain $I_j
\subset M$
\begin{eqnarray*}
    i_U d\,\TKK= d\,\SKK + \Lie_U\TKK,
\end{eqnarray*}
where
\begin{eqnarray}
    \label{CauchS} \SKK\equiv -i_U\TKK
\end{eqnarray}
is the total Cauchy stress 2-form\footnote{This follows from the
Cartan identity: $\Lie_{X}=i_{X}d + di_{X}$ for any $X \in \sect
TM$.}. If one identifies an electromagnetic $K\!$-drive $\TEM
{}^{I_j}$ in $\TKK$, such that
\begin{eqnarray*}
    \TKK=\TEM{}^{I_j} + \TKKK {}^{I_j}
\end{eqnarray*}
and
\begin{eqnarray*}
    \SKK=\SEM{}^{I_j} + \SKKK {}^{I_j},
\end{eqnarray*}
one then has
\begin{eqnarray*}
    \dot{P}_K^{mech\,U}{}^{I_j} [\Sig] + \dot{P}_K^{EM \,U}{}^{I_j} [\Sig] + \dot{P}_K^{rem \,U}{}^{I_j} [\Sig] = f_K^{EM\,\,U}{}^{I_j} [\Sig] + f_K^{rem\,\,U}{}^{I_j} [\Sig],
\end{eqnarray*}
where
\begin{eqnarray*}
    \dot{P}_K^{rem \,U}{}^{I_j} [\Sig] &\equiv& -\int_{\Sig} \Lie_U \TKKK {}^{I_j} \\
    \dot{P}_K^{EM \,U}{}^{I_j} [\Sig] &\equiv& -\int_{\Sig} \Lie_U \TEM{}^{I_j}
\end{eqnarray*}
denote integrated rates of change associated with field momenta in
$\TKKK {}^{I_j}$ and $\TEM{}^{I_j}$ respectively, and
\begin{eqnarray*}
    f_K^{rem\,\,U}  {}^{I_j} [\Sig] &\equiv& {\int_{\Sig}} d\,\SKKK {}^{I_j} \\
    f_K^{EM\,\,U}{}^{I_j} [\Sig] &\equiv& {\int_{\Sig}} d\,\SEM {}^{I_j}
\end{eqnarray*}
denote integrated forces associated with stresses in
$\TKKK{}^{I_j} $ and $\TEM{}^{I_j}$ respectively.\\

\section{The General Integrated Force Form in an Irregular Static Domain Composed of Different Media}
\label{mediaforce}
Suppose $\Sig=\sum_j \,I_j$ with $\SKK$ the Cauchy stress 2-form
for domain $I_j$ in a Minkowski spacetime with frame
$U=\frac{1}{\cc}\pdiff{}{t}$. In general, ${\TK}^{{field}}$ must
contain (time dependent) constraining forces to maintain the
overall equilibrium condition
\begin{eqnarray*}
    i_{\pdiff{}{t}}d\,\tau_K^{{field}}=0
\end{eqnarray*}
from stresses in each sub-domain $I_j$ of $\Sig$. In the (possibly
constrained) static case, $\Lie_U\TK=0$ and each 2-form
\begin{eqnarray*}
    i_{\pdiff{}{t}}d\,{\TKK}= d\SKK
\end{eqnarray*}
contributes an integrated reaction force on $\Sig$ from domain
$I_j$.

In general, each 4-velocity $\Vj \in \sect T I_{j}$ must be
determined from the jump conditions for $\TKK$. In the static
case, one has all $\nabla \Vj=0$ with $\Vj=U$, and one may define
the net integrated $K\!$-force for $\Sig$ in the frame $U$:
\begin{eqnarray*}
    f_K^{U\,}[\Sig]\equiv \sum_j \int_{\Sigma^3_{{I_j}}} i_{\pdiff{}{t} } d\,\tau_K^{{field}\,\, {I_j}} =   \sum_j \int_{\Sigma^3_{{I_j}}}  d\,\sigma_K^{U\,\,field{I_j}}  =   \sum_j \int_{\partial\Sigma^3_{{I_j}}}  \sigma_K^{U\,\,field{I_j}}.
\end{eqnarray*}
There may be additional sources of stress with support on
submanifolds of $M$. Singular sources of stress in the
electromagnetic field include charges, currents and their
multipoles, with support on points, lines $\Sigma^1$ or surfaces
$\Sigma^2$ in space \cite{RWT_JMP}. If the integrals on the right below are
finite, the most general integrated force can then be written so
as to include such distributional sources:
\begin{eqnarray*}
    f_K^{U}[\Sig]\equiv \sum_j \int_{\partial\Sigma^3_{{I_j}}}  \sigma_K^{U\,\,field{I_j}}+\sum_j \int_{\Sigma^2_{{I_j}}}  \kappa_K^{U\,\,field{I_j}}+ \sum_j \int_{\Sigma^1_{{I_j}}}  \gamma_K^{U\,\,field{I_j}},
\end{eqnarray*}
in terms of line stress 1-forms $\gamma_K$ and surface stress
2-forms $\kappa_K$.

A number of sources of interfacial stress depend on the local mean
curvature normal of the interface. For example, if the history of
the interface $\partial\,I_j$ is the spacetime hypersurface $f=0$
with unit spacelike normal ${N}= \frac{\widetilde{d\,f}}{\vert
{d\,f}\vert}$, then the scalar $(Tr\,\, H)$ is defined by
\begin{eqnarray*}
    d\,i_N i_U \star 1=(Tr\,\,H)\,\, i_U \star 1
\end{eqnarray*}
and $\eta \equiv (Tr\,\, H)\, N$ is the mean curvature normal.
{\it Surface tension} at an arbitrary interface depends on $\eta$
and the local surface tension scalar field $\gamma$, yielding the
particular interface forces:
\begin{eqnarray*}
    \int_{\Sigma^2_{{I_j}}  = \partial \Sigma^3_{{I_j}} }  \kappa_K^{U\,\,field{I_j}} &=& \int_{\Sigma^2_{{I_j}}} \left( \gamma \,i_K\,\wt{\eta}+ i_K\,d\,\gamma\right) i_N\,i_U\,\star 1 \\
    \int_{\Sigma^1_{{I_j}}}  \gamma_K^{U\,\,field{I_j}} &=&   \int_{ \Sigma^1_{{I_j}}= \partial \Sigma^2_{{I_j}}} \gamma \,i_N\,i_U\,i_K\,\star 1.
\end{eqnarray*}\\

\section{Electromagnetic Fields in Spacetime}
\label{ch_fields}
Maxwell's equations for an electromagnetic field in an arbitrary
medium can be written
\begin{eqnarray}\label{Maxwell}
    d\,F=0 \qquadand d\,\star\, G =j,
\end{eqnarray}
where $F\in\sect\Lambda^2 M$ is the Maxwell 2-form, $G\in
\sect\Lambda^2 M$ is the excitation 2-form and $j\in\sect\Lambda^3
M$ is the 3-form electric current source\footnote{All
  electromagnetic tensors in this article have dimensions
  constructed from the SI dimensions $[M], [L], [T], [Q]$ where $[Q]$
  has the unit of the Coulomb in the MKS system. We adopt $[g]=[L^2],
  [G]=[j]=[Q],\,[F]=[Q]/[\ep{0}]$ where the permittivity of free space
  $\ep{0}$ has the dimensions $ [ Q^2\,T^2 M^{-1}\,L^{-3}] $ and
  $c_0=\frac{1}{\sqrt{\ep{0}\mu_0}}$ denotes the speed of light in vacuo.
  Note that, with $[ g ]=[L^2]$, for $p$-forms $\alpha$ in $n$ dimensions
  one has $[\star \alpha]=[\alpha] [L^{n-2p}]$.}.
To close this system, ``electromagnetic constitutive relations''
relating $G$ and $j$ to $F$ are necessary. The functional tensor
relations
\begin{eqnarray*}
    G={\cal Z} [F]
\end{eqnarray*}
and
\begin{eqnarray*}
    j= {\cal Z}_1[F]
\end{eqnarray*}
are typical for idealized material without
electrostriction losses.

The electric 4-current $j$ describes both (mobile) electric charge
and effective (Ohmic) currents in a conducting medium. The {\it
electric field} $\Me\in\sect\Lambda^1 M$ and {\it magnetic
induction field} $\Mb\in\sect\Lambda^1 M$ associated with $F$ are
defined with respect to an observer field $U$ by
\begin{eqnarray}\label{intro_e_b}
    \Me = i_U F \qquadand \cc\Mb = i_U \star F.
\end{eqnarray}
Thus, $i_U\Me=i_U\Mb=0$ and with $g(U,U)=-1$,
\begin{eqnarray}\label{intro_F}
    F=\Me\w \wt{U} - \star\,(\cc\Mb\w \wt{U}).
\end{eqnarray}
Likewise the {\it displacement field} $\Md\in\sect\Lambda^1 M$ and
the {\it magnetic field} $\Mh\in\sect\Lambda^1 M$ associated with
$G$ are defined with respect to $U$ by
\begin{eqnarray}\label{Media_d_h}
    \Md = i_U G  \qquadand \frac{\Mh}{\cc} = i_U\star G.
\end{eqnarray}
Thus,
\begin{eqnarray}\label{Media_G}
    G=\Md\w \wt{U} - \star\,\(\frac{\Mh}{\cc}\w \wt{U}\),
\end{eqnarray}
with $i_U\Md=i_U\Mh=0$. The spatial 1-forms
$\Me,\,\Mb,\,\Md,\,\Mh$ are fields on a general spacetime defined
with respect to the frame $U$, which may be non-inertial if
$d\wt{U}\neq 0$.\\

\section{Time-Dependent Maxwell Systems in Space}
\label{ch time}
In the following, attention is restricted to fields on Minkowski
spacetime. This can be globally foliated by $3-$dimensional
spacelike hyperplanes. The Minkowski metric on spacetime induces a
metric with Euclidean signature on each spacetime hyperplane.
Furthermore, each hyperplane contains events that are deemed
simultaneous with respect to a clock attached to any integral
curve of a future-pointing, unit, time-like vector field
$U=\frac{1}{\cc}\frac{\partial}{\partial t}$ defining an inertial
observer on Minkowski spacetime and the spacetime Hodge map
$\star$ induces a Euclidean Hodge map $\#$ on each hyperplane by
the relation
\begin{eqnarray*}
    \star 1= \cc\, d t \w \# 1 \= \#1 \w \wt{U}.
\end{eqnarray*}
The spacetime Maxwell system can now be reduced to a family of
parameterised exterior systems on $\RR^3$. Each member is an
exterior system involving forms on $\RR^3$ depending
parametrically on some time coordinate $t$ associated with $U$.
Let the $(3+1)$ split of the 4-current 3-form with respect to the
foliation be
\begin{eqnarray*}
    j = - \bbfJ \w dt + \rho \# 1,
\end{eqnarray*}
with $i_{\pdiff{}{t}} \bbfJ =0 $. Then, from (\ref{Maxwell}),
\begin{eqnarray}\label{dj}
    d\, j=0
\end{eqnarray}
yields
\begin{eqnarray}\label{cont}
    \D \bbfJ + \dot{\rho} \# 1 = 0.
\end{eqnarray}
Here, and below, an over-dot denotes (Lie) differentiation with
respect to the parameter $t$ ($\dot\alpha\equiv
\Lie_{\pdiff{}{t}}\alpha$ for all $\alpha$) and $\D$ denotes
exterior differentiation on $\RR^3$ such that
\begin{eqnarray*}
    d &\equiv& \D + dt \w \Lie_{\pdiff{}{t}}.
\end{eqnarray*}
It is convenient to introduce on each spacetime hyperplane the
(Euclidean Hodge) dual forms:
\begin{eqnarray*}
    \bfE\equiv\# \Me,\quad && \quad \bfD\equiv\#\Md \\
    \bfB\equiv\# \Mb,\qquad \bfH&\equiv&\#\Mh,\qquad \bbfj\equiv\#\bbfJ,
\end{eqnarray*}
so that the $(3+1)$ split of the spacetime covariant Maxwell
equations (\ref{Maxwell}) with respect to $\wt{U}=-\cc dt$ becomes
\begin{eqnarray}
    \label{M1} \D\Me &=& -\dot{\bfB}, \\
    \label{M2} \D\bfB &=& 0, \\
    \label{M3} \D\Mh &=& \bbfJ  +\dot{\bfD}, \\
    \label{M4} \D\bfD &=& \rho\# 1.
\end{eqnarray}
All $p$-forms ($p\ge 0$) in these equations are independent of
$dt$, but have components that may depend parametrically on $t$.\\

\section{Electromagnetic Constitutive Tensors for Linear Media}
\label{linearmedia}
Attention will now be turned to integrated electromagnetic forces
on a class of polarisable media. This requires a discussion of a
class of electromagnetic constitutive tensors for linear media. In
general, the excitation tensor $G$ is a functional of the Maxwell
field tensor $F$ and properties of the medium
\begin{eqnarray*}
    G = {\cal Z}[F,\ldots].
\end{eqnarray*}
Such a functional induces, in general, non-linear and non-local
relations between $\Md, \Mh$ and $\Me, \Mb$. Electrostriction and
magnetostriction arise from the dependence of ${\cal Z}$ on the
elastic deformation tensor of the medium. For general {\it linear
continua}, one may define a collection of {\it constitutive tensor
fields} $Z^{\,(r)}$ on spacetime by the relation
\begin{eqnarray*}
    G = \Sigma_{r=0}^{N}{ Z^{\,{(r)}}}[\nabla^{\,r} F,\ldots],
\end{eqnarray*}
in terms of the spacetime connection (covariant derivative)
$\nabla$.

\def\eprr{\epsilon_r}

\def\murr{\mu_r}

In idealized (non-dispersive) {\it simple media}, one adopts the
simplified {\it local} relation
\begin{eqnarray*}
    G=Z(F),
\end{eqnarray*}
for some {degree 4 constitutive tensor field} $Z$ and in the
vacuum $G=\epsilon_0 F$. Regular {\it linear isotropic media} are
described by a bulk 4-velocity field $V$, a relative permittivity
scalar field $\eprr{}$ and a non-vanishing relative permeability
scalar field $\murr$. In this case, the structure of $Z$ follows
from
\begin{eqnarray*}
    \frac{G}{\ep{0}} = \eprr{} \,i_V F \w \wt{V} - \murr^{-1}\star(i_V\star F\w\wt{V}) \= \( \eprr{} - \frac{1}{\murr} \) \, i_V F \w \wt{V} + \frac{1}{\murr} \, F.
\end{eqnarray*}
In a comoving frame with $U=V$, this becomes
\begin{eqnarray*}
    \Md = \ep{0}\eprr{}\, \Me \qquadand \Mh = (\mu_0\murr)^{-1} \Mb.
\end{eqnarray*}
To discuss linear (non-dispersive, lossless), inhomogeneous, {\it
anisotropic media}, it is convenient to describe $Z$ in a
particular basis associated with the medium. Since $Z$ is a tensor
that maps 2-forms to 2-forms, in any spacetime local frame
$\{e^0,e^1,e^2,e^3\}$, one may write
\begin{eqnarray*}
    \tfrac12 G_{ab}e^a \w e^b = \tfrac14 Z^{c d}{}_{a b} F_{c d} e^a \w e^b,
\end{eqnarray*}
where
\begin{eqnarray*}
    Z^{c d}{}_{a b} = - Z^{c d}{}_{b a} = - Z^{d c}{}_{a b} = Z^{d c}{}_{b a}.
\end{eqnarray*}
Thus, $Z$ can be described in terms of spatial rank $3$ tensors on
spacetime, relating observed electric and magnetic fields in some
frame $U$, with:
\begin{eqnarray*}
    \Md &=& \PermDE(\Me) + \PermDB(\Mb) \\
    \Mh &=& \PermHE(\Me) + \PermHB(\Mb).
\end{eqnarray*}
In such a frame, the medium is said to exhibit {\it
magneto-electric} properties in general. If $\PermDB$ and
$\PermHE$ are non-zero in the co-moving frame of the medium, it is
called magnetoelectric. If $\PermDB$ and $\PermHE $ are zero in
the co-moving frame of the medium, it is called
non-magnetoelectric. The spatial tensors $\PermDB$ and $\PermHE$
may be non-zero in a non-comoving frame for a non-magnetoelectric
medium. Due to the behaviour of electric and magnetic fields under Lorentz transformations, all materials exhibit magnetoelectric properties in some
frame. Thermodynamic and time symmetry conditions impose the
relation $Z=Z^{\dagger}$ \cite{ODell} or
\begin{eqnarray*}
    \PermDE^\dagger=\PermDE \,,\quad \PermHB^\dagger=\PermHB \quad\text{and}\quad \PermDB^\dagger=-\PermHE
\end{eqnarray*}
in all spacetime frames, where the adjoint $T^\dagger$ of a tensor
$T$ which maps $p$-forms to $p$-forms is defined by:
\begin{eqnarray*}
    \alpha\w\star T(\beta) = \beta\w\star T^\dagger(\alpha) \qquad\text{for all }\alpha,\beta\in\sect\Lambda^p M.
\end{eqnarray*}\\

\section{Homogeneous Dispersive Magnetoelectric Media}
\label{SEM}
In dispersive media, constitutive relations between the spatial
fields $\Me,\,\Mb,\,\Md,\,\Mh$ are non-local in spacetime. If the
medium is {\it spatially homogenous}, so that it has no preferred
spatial origin, then it is possible to Fourier transform the
fields with respect to space and time, and work with transformed
local constitutive relations.

For any real valued $p$-form $\alpha$, define its complex valued
Fourier transform $\FA$ by
\begin{eqnarray}\label{FT}
    \alpha=\int_{-\infty}^\infty \,d\omega \int_{-\infty}^\infty d\kk\,\,\FA \,\exp{i(\kk \cdot \rr - \omega t)},
\end{eqnarray}
where $\kk\in\RR^3$. Then the source free Maxwell
system reduces to
\begin{eqnarray}
    \label{F1} \KKK \w \Fee{}{} &=& \omega\FBB{}{} \\
    \label{F2} \KKK \w \Fhh{}{} &=& -\omega\FDD{}{},
\end{eqnarray}
where the real propagation wave 1-form $\KKK\equiv\kk\cdot d\rr
\in \sect\Lambda^1 M$. The remaining transformed Maxwell equations
$\KKK\w \FBB{}{}=0$ and $\KKK\w \FDD{}{}=0$ follow trivially from
(\ref{F1}) and (\ref{F2}) when $\omega\neq 0$. It also follows trivially that $
\Fee{}{}\w \FBB{}{} = 0\; (\mbox{ i.e.  } \Fee{}{} \mbox{ is
perpendicular to } \Fbb{}{})$. Similarly, $\FBB{}{}\w\KKK=0$ and
$\FDD{}{}\w\KKK=0$.

We assume that the magnetoelectric constitutive relations take the
form
\begin{eqnarray}
    \label{FCR1} \Fdd{}{} &=& \Xde(\Fee{}{})+\Xdb(\Fbb{}{}) \\
    \label{FCR2} \Fhh{}{} &=& \Xhe(\Fee{}{})+\Xhb(\Fbb{}{}).
\end{eqnarray}
These will (by convolution) give rise to non-local spacetime
constitutive relations. We also maintain the above symmetry
properties on the magnetoelectric tensors $\Xde, \Xdb, \Xhe,
\Xhb$. Substituting (\ref{FCR1}) and (\ref{FCR2}) in (\ref{F1})
and (\ref{F2}) yields a degenerate 1-form linear eigen-equation
for $\Fee{}{}$:
\begin{eqnarray}
    \nonumber \omega^2 \Xde(\Fee{}{}) + \omega \Xdb  \( \#( \KKK \w \Fee{}{} ) \) &+& \omega \# \( \KKK \w \Xhe(\Fee{}{}) \)  \\ \label{FDR} &+& \# \(\KKK\w \Xhb\(\#(\KKK\w\Fee{}{}) \) \) = 0.
\end{eqnarray}
The field $\Fbb{}{}$ then follows from (\ref{F1}), (up to a
scaling) and $\Fdd{}{},\Fhh{}{}$ from (\ref{FCR1}),(\ref{FCR2})
respectively. Equation (\ref{FDR}) may be written
\begin{eqnarray}\label{FD1}
    \XDD(\Fee{}{})=0,
\end{eqnarray}
defining the $1-1$ tensor $\XDD$. For non-trivial solutions
$\Fee{}{}$, the determinant of the matrix $\XD$ representing
$\XDD$ must vanish:
\begin{eqnarray}\label{det}
    det(\XD)=0.
\end{eqnarray}
Note that, in general, the roots of this dispersion relation are
not invariant under the transformation $\KKK\to\,-\KKK$. If one
writes $\kk= \hat\kk |\kk|$ in terms of the Euclidean norm
$|\kk|$, and introduces the refractive index ${\cal N}=|\kk|
\frac{\cc}{\omega}>0$ and $\hat\kk$ in place of $\kk$, then
solutions propagating in the direction described by ${\hat\kk}$
with angular frequency $\omega>0$ correspond to roots of
(\ref{det}) (labelled $r$) that may be expressed in the form
${\cal N}_r={\cal F}_r(\hat\kk,\omega)$. Thus, there can be a set
of distinct characteristic waves each with its unique refractive
index that depends on the propagation direction $\hat\kk$ and
frequency $\omega$. When the characteristic equation (\ref{det})
is a quadratic polynomial in ${\cal N}^2$ and has two distinct
roots that describe two distinct propagating modes for a given
$\omega$, the medium is termed {\it birefringent}. Roots ${\cal
N}^2_r$ such that ${\cal N}_r(\hat\kk, \omega) \neq {\cal
N}_r(-\hat\kk,\omega)$ imply that harmonic plane waves propagating
in the opposite directions $\pm\hat\kk$ have different wave
speeds.

Each eigen-wave will have a uniquely defined polarisation obtained
by solving the independent equations in (\ref{FD1}) for
$\Fee{r}{}$, up to normalisation. Since $\Fee{r}{}$ is complex, it
is convenient to introduce the eigen-wave normalisation by writing
\begin{eqnarray*}
    \Fee{r}{} ={\FFee}^r\, \Fnnr,
\end{eqnarray*}
in terms of the complex 0-form ${\FFee}^r$ and complex
polarisation 1-form ${\Fnn}^r$, normalised to satisfy
\begin{eqnarray}
    \conj{\Fnnr}\w \#{\Fnnr} =\#\,1
\end{eqnarray}
for each $r$. If one applies $\#\,\overline{\Fee{r}{}}\w\#$ to
(\ref{FDR}), making use of the symmetries between the real
magnetoelectric tensors $\Xde,\,\Xdb,\,\Xhe,\,\Xhb$, and evaluates
it with the eigen-wave $\Fee{r}{}$, one obtains the {\it real}
0-form dispersion relation for the characteristic mode $r$:
\begin{eqnarray*}
    \omega^2  \# \( \conj{\Fnnr}  \w \# \Xde(\Fnnr) \) &+& \omega\, \#\( \conj{\Fnnr} \w \#\Xdb \( \#\,(\KKK \w  \Fnnr) \)\)   \\
    +   \omega \,\# \( \conj{\Fnnr} \w \KKK \w \Xhe(\Fnnr) \)  &+& \# \(\conj{\Fnnr} \w \KKK \w \Xhb \( \#(\KKK \w \Fnnr) \)\) = 0,
\end{eqnarray*}
where $\KKK=\frac{\omega}{\cc} {\cal N} \hat\kk\cdot d\rr$ in
terms of ${\cal N}$ and $\hat\kk$.\\

\section{Electromagnetic Stress-Energy-Momentum Tensors}
\label{SEM}
There has been intense debate over many decades about the
appropriate choice of electromagnetic stress-energy-momentum
tensor that transmits forces in a (moving) polarisable medium
\cite{Pfeifer}. In 1909, Abraham introduced the {{\it symmetric}}
electromagnetic stress-energy-momentum tensor $\TT^{EM}$ for a
medium with 4-velocity $V$:
\begin{eqnarray*}
    2\,\TT^{EM}= -i_a F\otimes i^a G - i_a G \otimes i^a F - \star(F\w\star G) g + \wt{V} \otimes \GenPoyn + \GenPoyn \otimes\wt{V},
\end{eqnarray*}
where $i_a\equiv i_{X_a},\,\, i^a\equiv g^{ab}\,i_b$ in any vector
basis $\{X_{a}\}$ and
\begin{eqnarray*}
    \GenPoyn &=& \star\(\frac{1}{\cc}\Me^V\w\Mh^V\w\wt{V} - \cc\Md^V\w\Mb^V\w\wt{V}\),\\
    \text{where} \qquad \Me^V &=& i_V F, \qquad \cc\Mb^V \= i_{V}\star F, \qquad \Md^V \= i_{V}G  \qquadand \frac{\Mh^V}{\cc} \= i_V\star G,
\end{eqnarray*}
are fields defined relative to the motion of the medium, so that
\begin{eqnarray*}
    F &=& \Me^V\w \wt{V} - \star\,\(\cc\Mb^V \w \wt{V}\) \\
    G &=& \Md^V\w \wt{V} - \star\,\(\frac{\Mh^V}{\cc} \w \wt{V}\)
\end{eqnarray*}
with
\begin{eqnarray*}
    G = Z(F).
\end{eqnarray*}
For any Killing field $K$ the drive form associated with Abraham's electromagnetic
stress-energy-momentum tensor is
\begin{eqnarray}
    \label{AbrSF} \tau^{EM}_{K} = \frac{1}{2}\(F\w i_{K}\star G - i_{K}G \w \star F + \GenPoyn(K)\star \wt{V} + \wt{V}(K)\star\GenPoyn\).
\end{eqnarray}
It follows from (\ref{CauchS}), (\ref{intro_F}) and
(\ref{Media_G}) that
\begin{eqnarray}
    \label{AbrStress} \qquad J^{U}_{K}\equiv\sigma^{U}_{K} &=& \frac{1}{2}\( \df \Me(K)\#\Md+ \Md(K) \# \Me + \Mh(K)\# \Mb +\Mb(K)\#\Mh \) \\
    \nonumber                                           & & - \frac{1}{2}\#\( \df \Me \w \# \Md +  \Mb \w \#\Mh \)\#\wt{K}  + \dfrac{1}{2}\wt{U}(K)\( \dfrac{1}{\cc}\Me \w \Mh + \cc\Md \w \Mb \) \\
    \nonumber && + \dfrac{1}{2}i_{U}\(\wt{K} \w i_{V}\star s \)  - \wt{V}(K)i_{U}\star s
\end{eqnarray}
and
\begin{eqnarray}
    \label{AbrD} \rho^{U}_{K} &=& -\dfrac{1}{2}\wt{U}(K)\( \Mb \w \# \Mh + \Me\w \# \Md\) + \frac{1}{2}\( \dfrac{1}{\cc}\Me \w \Mh+\cc\Md \w \Mb\)\w \wt{K}^{\perp} \\ \nonumber & & -\frac{1}{2}i_{U}\( \wt{K} \w \wt{U} \w i_{V}\star s \) + \wt{V}(K)\,i_{U}(\star\, s\w\wt{U}),
\end{eqnarray}
where $K^{\perp}\equiv K + \wt{U}(K)U$.

By contrast, Minkowski (1908)
introduced the {{\it non-symmetric}} electromagnetic
stress-energy-momentum tensor $\TT^{EM}$ where
\begin{eqnarray}
   \label{MinkSEM} \TT^{EM}= -i_a F\otimes i^a G - \frac{1}{2}\star(F\w\star\, G) g,
\end{eqnarray}
which exhibits no explicit dependence on the medium 4-velocity
$V$. The corresponding drive form is
\begin{eqnarray*}
    \label{MinkSF} \tau^{EM}_{K} = \frac{1}{2}\(F\w i_{K}\star G - i_{K}F \w \star G\)
\end{eqnarray*}
and (\ref{CauchS}), (\ref{intro_F}) and (\ref{Media_G}) yield in this case
\begin{eqnarray}
    \label{MinkStress} J^{U}_{K}\equiv\sigma^{U}_{K} &=& \Mh(K)\# \Mb + \Me(K)\#\Md + \frac{1}{\cc}\wt{U}(K)\Me \w \Mh \\
    \nonumber                                        & & - \frac{1}{2}\#\( \Me \w \# \Md + \Mb \w \#\Mh\)\#\wt{K}.
\end{eqnarray}
and
\begin{eqnarray}
    \label{MinkD} \rho^{U}_{K} &=& \cc\Md \w \Mb \w \wt{K}^{\perp} - \frac{1}{2}\wt{U}(K)\( \Me \w \# \Md + \Mb \w \# \Mh\).
\end{eqnarray}

More recently other choices for an  electromagnetic
stress-energy-momentum tensor have been proposed which in themselves simply imply different constitutive relations \cite{Maugin1} with respect to a particular {\it total} stress-energy-momentum tensor.
In \cite{DGT}, \cite{DGT2}, it has been argued that different
choices of the electromagnetic stress-energy-momentum tensor for linear
polarisable media are equivalent to different choices of $Z$ {\it
and} a different partition of the {\it total}
stress-energy-momentum tensor for the computation of so called
pondermotive forces that arise from the divergence of terms in its
decomposition. Furthermore, it was shown how particular choices of
the dependence of $Z$ on the gravitational interaction led, via a covariant
variational formulation, to either the Abraham tensor or a
symmetrized version of that proposed by Minkowski.

In the following, we illustrate how the general theory of drive
forms outlined above offers a natural tool to discuss the
computation of particular electromagnetic forces for materials
that exhibit magnetoelectric properties (at rest) in the
laboratory, for a particular choice of electromagnetic drive form.
This is an essential step in any program that attempts to confront
experimental measurements of such forces with theoretical
prediction.

To facilitate this calculation, an {\it electromagnetic} drive form
associated with the tensor obtained by symmetrizing
(\ref{MinkSEM}) will be chosen:
\begin{eqnarray}
    \label{SymMinkEM} {\tau_K}^{EM}=\frac{1}{2}(F\w i_K \star G - i_K G \w \star F).
\end{eqnarray}
It follows from (\ref{CauchS}), (\ref{intro_F}) and
(\ref{Media_G}) that with this drive-form:
\begin{eqnarray}
    \label{SymMinkStress} \qquad J^{U}_{K}\equiv\sigma^{U}_{K} &=& \frac{1}{2}\( \df \Me(K)\#\Md+ \Md(K) \# \Me + \Mh(K)\# \Mb +\Mb(K)\#\Mh \) \\
    \nonumber                                           & & - \frac{1}{2}\#\( \df \Me \w \# \Md +  \Mb \w \#\Mh \)\#\wt{K}  + \frac{1}{2}\wt{U}(K)\( \frac{1}{\cc}\Me \w \Mh + \cc\Md \w \Mb \).
\end{eqnarray}
and
\begin{eqnarray}
    \label{SymMinkD} \rho^{U}_{K} &=& \frac{1}{2}\(\frac{1}{\cc}\Me \w \Mh + \cc\Md \w \Mb \)\w \wt{K}^{\perp} - \frac{1}{2}\wt{U}(K)\( \Me \w \# \Md  + \Mb \w \# \Mh\).
\end{eqnarray}
For a medium at rest in the laboratory,
$U=V=\frac{1}{\cc}\partial_{t}$. Furthermore, if $\wt{U}(K)=0$,
the 2-forms (\ref{AbrStress}) and (\ref{SymMinkStress}) coincide, so the following analysis does not discriminate between the choice of tensors (\ref{AbrSF}) and (\ref{SymMinkEM}). However, in this case the instantaneous densities (\ref{AbrD}) and (\ref{SymMinkD}) are
different. But, for the polarised monochromatic plane waves discussed below, the time-averaged tensors based on (\ref{MinkD}) and (\ref{SymMinkD}) also coincide.

If the fields are all differentiable in the medium described by (\ref{SymMinkEM}), one readily
obtains
\begin{eqnarray*}
    d\,\tau_K^{EM} &=& \frac{1}{2}\left(i_K dG \w \star F + i_KG \w d\star F - F \w i_K d\star G \right) \\
                   &=& F \w \star i_K d\(\,\frac{\Pi}{2}\) - \left(F + \left(\frac{\Pi}{2\ep{0}}\right)\right)\w \,i_K \,j +G\w i_K\,d\star\left( \frac{\Pi}{2\ep{0}}\right),
\end{eqnarray*}
where
\begin{eqnarray*}
    d\,F=0,\qquad  d\,\star G=j,\qquad G=\ep{0}F+\Pi,\qquad \ep{0}\,d\,\star F = {j}-d\,\star\Pi.
\end{eqnarray*}
Thus, non-zero bulk integrated static electromagnetic forces from
such fields require $d\Pi\neq 0, d\star\Pi\neq 0$ (magnetisation
or electrical polarisation inhomogeneities) or $j\neq 0$ (non-zero
local source current or charge density). For a neutral homogeneous
material therefore, we consider a medium whose electromagnetic
properties change discontinuously at some interface.


\section{The Magnetoelectric Slab}
\label{MEslab}
In terms of the rank 3 identity tensor $\Id$ in space, consider an
infinitely extended slab\footnote{Such a medium has been
considered by Hehl and Obukhov in their classical analysis of the
Feigel effect \cite{HO}.} of magnetoelectric material with
\begin{eqnarray}
    \label{XdeMS} \Xde &=& \ep{\Y} \,\, \Id \\
    \label{XhbMS} \Xhb &=& {\mu}^{-1}_{\Y}\,\,  \Id.
\end{eqnarray}
The slab has width $L$ and parallel interfaces (with the vacuum) at $x=0$ and $x=L$. It is oriented in the
laboratory frame $\{\partial_x,\partial_y,\partial_z\}$, so that
$\Xdb$ takes the particular form
\begin{eqnarray}
    \label{XdbMS} \Xdb = {\beta_{1, {\boldsymbol k},\omega}} \,\,dz \otimes  \partial_y+ {\beta_{2, {\boldsymbol k},\omega}} \,\, dy \otimes \partial_z,
\end{eqnarray}

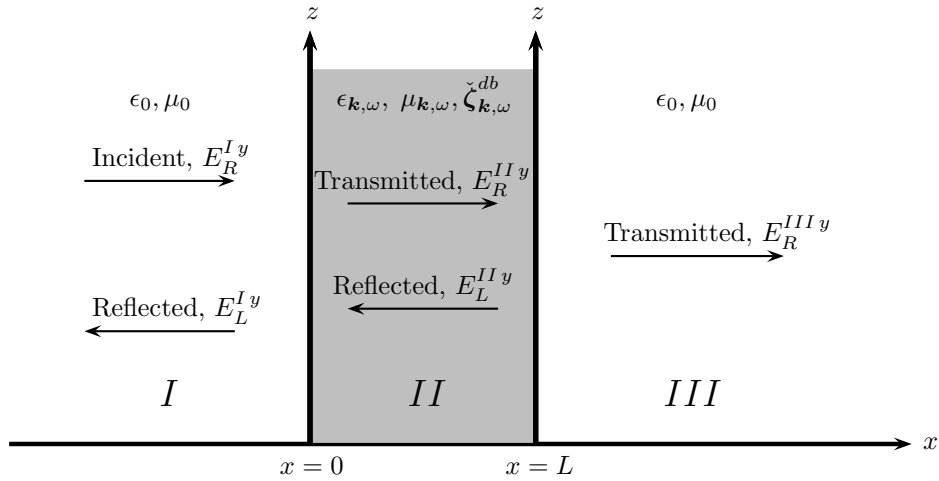
\begin{figure}[h]
\setlength{\unitlength}{1cm}
\begin{center}
\begin{picture}(10,7)
    \put(2.94,6.66){$z$}
    \put(5.94,6.66){$z$}
    \psline[linewidth=1.7pt]{->}(-1,1)(11,1)
    \put(11.15,0.94){$x$}
    \psframe[linestyle=none,fillstyle=solid,fillcolor=lightgray](3,1)(6,6)

    \psline[linewidth=2pt]{->}(6,1)(6,6.5)
    \psline[linewidth=2pt]{->}(3,1)(3,6.5)

    \psline[arrowsize=3pt 2]{->}(0,4.5)(2,4.5)
    \psline[arrowsize=3pt 2]{->}(3.5,4.2)(5.5,4.2)
    \psline[arrowsize=3pt 2]{->}(2,2.5)(0,2.5)
    \psline[arrowsize=3pt 2]{->}(5.5,2.8)(3.5,2.8)
    \psline[arrowsize=3pt 2]{->}(7,3.5)(9.3,3.5)
    \psline[linewidth=2pt]{->}(6,1)(6,6.5)
    \psline[linewidth=2pt]{->}(3,1)(3,6.5)
    \put(0.1,4.7){\text{Incident, $\FE{I}{R}{y}$}}
    \put(0.1,2.7){\text{Reflected, $\FE{I}{L}{y}$}}
    \put(3.05,4.4){\text{Transmitted, $\FE{II}{R}{y}$}}
    \put(3.3,3){\text{Reflected, $\FE{II}{L}{y}$}}
    \put(6.9,3.7){\text{Transmitted, $\FE{III}{R}{y}$}}

    \put(2.6,0.6){$x=0$}
    \put(5.6,0.6){$x=L$}
    \put(1,1.5){$\mbox{\Large $I$}$}
    \put(4.3,1.5){$\mbox{\Large $II$}$}
    \put(7.7,1.5){$\mbox{\Large $III$}$}

    \put(0.6,5.5){$\ep{0},\mu_{0}$}
    \put(7.6,5.5){$\ep{0},\mu_{0}$}
    \put(3.35,5.5){$\ep{\Y},\;\mu_{\Y},\Xdb$}
\end{picture}
\caption{Geometry of the magnetoelectric slab and the electric field amplitudes in the three regions.}
\end{center}
\label{MagSlabDiag}
\end{figure}
In this frame, the modes associated with the branch (\ref{polz})
of the dispersion relation below will be polarised in the direction
$\partial_z$ and those associated with the branch (\ref{poly})
will be polarised in the direction $\partial_y$. The matrix
representing $\Xdb$ takes the form
\begin{eqnarray}\label{Xdbmatrix}
    [\Xdb] \equiv \( \begin{array}{ccc}
                                            0 & 0 & 0 \\
                                            0 & 0 & {\beta_{1, {\boldsymbol k},\omega}}  \\
                                            0 & {\beta_{2, {\boldsymbol k},\omega}}  & 0 \\
                                        \end{array}\).
\end{eqnarray}
It follows that
\begin{eqnarray}
    \label{XheMS} \Xhe= -{\beta_{2, {\boldsymbol k},\omega}}  \,\,dz \otimes  \partial_y - {\beta_{1, {\boldsymbol k},\omega}} \,\, dy \otimes \partial_z.
\end{eqnarray}
With this choice of orientation of the slab, the spatial region
$0<x<L$ will be denoted $II$ and the region with $x>L$ denoted
$III$. The electromagnetic fields induced in its interior by a
plane monochromatic wave normally incident
from the right (in region $I$, $x< 0$) propagating in the
direction $\partial_x$ with polarisation in the direction
$\partial_y$ can now be readily determined.

From (\ref{det}), the dispersion relation associated with one
polarised eigen-mode of $\Fee{}{}$ is
\begin{eqnarray}\label{polz}
    \ep{\Y}\mu_\Y \omega^2 - k^2 - 2{\beta_{1, {\boldsymbol k},\omega}}  k \omega=0,
\end{eqnarray}
while that associated with the other polarised eigen-mode of
$\Fee{}{}$ is
\begin{eqnarray}\label{poly}
    \ep{\Y}\mu_\Y \omega^2 - k^2 + 2{\beta_{2, {\boldsymbol k},\omega}}  k \omega =0.
\end{eqnarray}
Each relation can describe propagating modes with angular
frequency $\omega>0$ moving in a direction determined by
$sign(k)\,\partial_x$ with phase speed $|\omega/k|$. Since this
ratio depends on the values of ${\beta_{1, {\boldsymbol
k},\omega}}$ or ${\beta_{2, {\boldsymbol k},\omega}}$, it may
exceed the speed of light in vacuo. In principle, such modes can
contribute to the synthesis of wave packets. However, in the
following, we restrict to monochromatic incident waves and work
with constitutive parameters that inhibit super-luminal waves,
with real constants $\ep{\Y}\equiv\ep{} >0, \,\mu_\Y\equiv\mu >0
,\, {\beta_{1, {\boldsymbol k},\omega}} \equiv\beta_1,  \,
{\beta_{2, {\boldsymbol k},\omega}}\equiv\beta_2 $. For an
incident wave with complex amplitude ${\cal E}$, no loss of
generality arises by taking $\omega>0$ and writing the solution
$\Fee{}{}$:
\begin{eqnarray}
    \label{FeeI} \Fee{I}{y} &=& {\cal E}\( \exp(i \Fk{I}{R}{y}  x -i\omega\, t ) \, dy + \FE{I}{L}{y} \,\exp(i\Fk{I}{L}{y} x-i\omega\, t)\,dy\) \\
    \label{FeeII} \Fee{II}{y} &=& {\cal E}\(\FE{II}{R}{y} \exp(i\Fk{II}{R}{y} x -i\omega\, t)\,dy + \FE{II}{L}{y} \exp({i \Fk{II}{L}{y} x-i\omega\, t  })\, dy\) \\
    \label{FeeIII} \Fee{III}{y} &=& {\cal E}\FE{III}{R}{y} \exp(i \Fk{III}{R}{y} x-i\omega\, t)\, dy
\end{eqnarray}
where $\Fk{II}{R}{y}$ denotes a real root of the dispersion
relation (\ref{poly}) associated with the polarisation eigenvector
$\partial_y$ with $sign(\Fk{II}{R}{y})> 1$, describing a polarised
right-moving wave in the slab (region $II$). Similarly,
$\Fk{II}{L}{y}$ denotes a real root of the dispersion relation
associated with the polarisation eigenvector $\partial_y$ with
$sign(\Fk{II}{L}{y})< 1$, describing a polarised left-moving wave
in the slab (region $II$). In general, these wave numbers are
different. In the vacuum regions, $\Fk{I}{R}{y} = -\Fk{I}{L}{y} =
\Fk{III}{R}{y} = \omega/\cc$.

If $\Omega_0^{*}$ ($\Omega_L^{*})$ denotes the pull-back of forms
to the interface $x=0$ ($x=L$), the interface boundary conditions
\cite{Jackson} are
\begin{eqnarray}
    \Omega_0^{*} \(\Fee{I}{y} - \Fee{II}{y}\) &=& \Omega_L^{*} \(\Fee{II}{y}-\Fee{III}{y}\) = 0  \\
    \Omega_0^{*} \(\Fhh{I}{y} - \Fhh{II}{y}\) &=& \Omega_L^{*} \(\Fhh{II}{y}-\Fhh{III}{y}\) = 0  \\
    \Omega_0^{*} \(\FBB{I}{y} - \FBB{II}{y}\) &=& \Omega_L^{*} \(\FBB{II}{y}-\FBB{III}{y}\) = 0  \\
    \Omega_0^{*} \(\FDD{I}{y} - \FDD{II}{y}\) &=& \Omega_L^{*} \(\FDD{II}{y}-\FDD{III}{y}\) = 0,
\end{eqnarray}
yielding the linear system for the dimensionless complex amplitudes $
\FE{I}{L}{y} \,, \FE{II}{R}{y} \,, \FE{II}{L}{y} \,,
\FE{III}{R}{y}$:
\begin{eqnarray*}
    1 + \FE{I}{L}{y} &=& \FE{II}{R}{y} + \FE{II}{L}{y} \\
    \(\frac{\Fk{II}{R}{y}}{\mu\omega}-\beta_{2} \)\FE{II}{R}{y} + \(\frac{\Fk{II}{L}{y}}{\mu\omega}-\beta_{2} \)\FE{II}{L}{y} &=& \frac{1}{\mu_{0}\omega}\(\df \Fk{I}{R}{y} + \FE{I}{L}{y}\Fk{I}{L}{y} \)  \\
    \FE{II}{R}{y}\exp({i\Fk{II}{R}{y}L}) + \FE{II}{L}{y}\exp({i\Fk{II}{L}{y}L}) &=& \FE{III}{R}{y}\exp({i\Fk{III}{R}{}L}) \\
    \(\frac{\Fk{II}{R}{y}}{\mu\omega}-\beta_{2} \)\FE{II}{R}{y}\exp({i\Fk{II}{R}{y}L}) + \(\frac{\Fk{II}{L}{y}}{\mu\omega}-\beta_{2} \)\FE{II}{L}{y}\exp({i\Fk{II}{L}{y}L}) &=& \frac{\Fk{III}{R}{y}\FE{III}{R}{y}}{\mu_{0}\omega}\exp({i\Fk{III}{R}{y}L}).
\end{eqnarray*}
This system of equations has the solution:
\begin{eqnarray}
\begin{split}\label{fieldsols}
    \quad \FE{I}{L}{y} &\= \frac{\Gamma_{-}^{\,y}}{\Gamma_{+}^{\,y}} \\
    \quad \FE{II}{L}{y} &\= \frac{\mu\(\Fk{I}{L}{y}-\Fk{I}{R}{y}\)\(\mu\mu_{0}\beta_{2}\omega + \Fk{III}{R}{}\mu-\mu_{0}\Fk{II}{R}{y}\)\exp({i\Fk{II}{R}{y}L})}{\Gamma_{+}^{\,y}} \\
    \quad \FE{II}{R}{y} &\= \frac{\mu\(\Fk{I}{R}{y}-\Fk{I}{L}{y}\)\(\mu\mu_{0}\beta_{2}\omega + \Fk{III}{R}{}\mu-\mu_{0}\Fk{II}{L}{y}\)\exp({i\Fk{II}{L}{y}L})}{\Gamma_{+}^{\,y}} \\
    \quad \FE{III}{R}{y} &\= \frac{\mu\mu_{0}\(\Fk{I}{R}{y}-\Fk{I}{L}{y}\)\(\Fk{II}{R}{y}-\Fk{II}{L}{y}\)\exp({i(\Fk{II}{R}{y}+\Fk{II}{L}{y}-\Fk{III}{R}{})L})}{\Gamma_{+}^{\,y}},
\end{split}
\end{eqnarray}
where it is convenient to introduce
\begin{eqnarray*}
    \Gamma_{\pm}^{\,y} &=& \( \exp({i\Fk{II}{R}{y}L}) - \exp({i\Fk{II}{L}{y}L})\df\)\[\Fk{I}{\pm}{y}\mu^{2}\(\mu_{0}\beta_{2}\omega + \Fk{III}{R}{y} \) \pm \mu\mu_{0}^{2}\beta_{2}\omega\( \beta_{2}\mu\omega  - \Fk{II}{R}{y} - \Fk{II}{L}{y} \)  \right. \\ & & \left. + \mu_{0}\( \beta_{2}\omega\mu^{2}\Fk{III}{R}{y} + \mu_{0}\Fk{II}{R}{y}\Fk{II}{L}{y}\)  \] \pm  \(\df \Fk{II}{R}{y}\exp({i\Fk{II}{L}{y}L}) - \Fk{II}{L}{y}\exp({i\Fk{II}{R}{y}L}) \)\mu\mu_{0}\Fk{III}{R}{y} \\ && + \(\df \Fk{II}{L}{y}\exp({i\Fk{II}{L}{y}L}) - \Fk{II}{L}{y}\exp({i\Fk{II}{L}{y}L}) \)\mu\mu_{0}\Fk{I}{\pm}{y}
\end{eqnarray*}
with
\begin{eqnarray*}
    \Fk{I}{+}{y} = \Fk{I}{R}{y} \qquadand \Fk{I}{-}{y} = \Fk{I}{L}{y}.
\end{eqnarray*}
With the electric field amplitudes determined, the complete set of
polarised fields\newline
$\{\Fee{}{y},\Fbb{}{y},\Fdd{}{y},\Fhh{}{y}\}$ in each region is
determined. For completeness, these fields are
given in the appendix.\\

\section{Average Pressure on the Magnetoelectric Slab}
\label{pressure}
To calculate the average pressure on the sides of the
magnetoelectric slab, one integrates the Maxwell-Cauchy stress
2-form over the 2-chain (surface) $\Omega=
\Omega_{0}+\Omega_{1}+\Omega_{L}+\Omega_{2}$ indicated
schematically in figure~(3). The image of $\Omega$ is the boundary
of a box of height $H$, width $W$ and length $L$, with faces
$\Omega_{0}$ and $\Omega_{L}$ in regions $I$ and $III$
respectively, parallel to the surfaces of the slab. Integrating
over a box with faces wholly within $II$ would give zero total
force, since region $II$ is homogeneous. Since the fields are
independent of $z$, contributions to the integral from the
oriented chains $\Omega_{1}$ and $\Omega_{2}$ cancel.

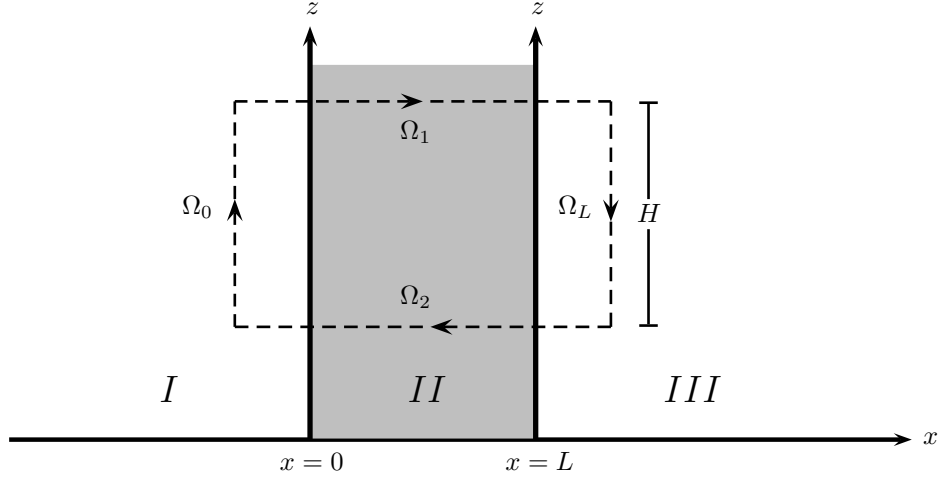
\begin{figure}[h]
\setlength{\unitlength}{1cm}
\begin{center}
\begin{picture}(10,7)
    \put(2.94,6.66){$z$}
    \put(5.94,6.66){$z$}
    \psline[linewidth=1.7pt]{->}(-1,1)(11,1)
    \put(11.15,0.94){$x$}
    \psframe[linestyle=none,fillstyle=solid,fillcolor=lightgray](3,1)(6,6)

    \psline[linewidth=2pt]{->}(6,1)(6,6.5)
    \psline[linewidth=2pt]{->}(3,1)(3,6.5)
    \psline[linewidth=1pt,linestyle=dashed,arrowsize=3pt 3]{->}(2,5.5)(4.5,5.5)
    \psline[linewidth=1pt,linestyle=dashed,arrowsize=3pt 3](4.6,5.5)(7,5.5)
    \psline[linewidth=1pt,linestyle=dashed,arrowsize=3pt 3]{->}(2,2.5)(2,4.2)
    \psline[linewidth=1pt,linestyle=dashed,arrowsize=3pt 3](2,4.2)(2,5.5)
    \psline[linewidth=1pt,linestyle=dashed,arrowsize=3pt 3](7,2.5)(7,3.9)
    \psline[linewidth=1pt,linestyle=dashed,arrowsize=3pt 3]{<-}(7,3.9)(7,5.5)
    \psline[linewidth=1pt,linestyle=dashed,arrowsize=3pt 3](2,2.5)(4.5,2.5)
    \psline[linewidth=1pt,linestyle=dashed,arrowsize=3pt 3]{<-}(4.6,2.5)(7,2.5)
    \put(1.3,4){$\Omega_{0}$}
    \put(6.3,4){$\Omega_{L}$}
    \put(4.2,5.0){$\Omega_{1}$}
    \put(4.2,2.8){$\Omega_{2}$}
    \psline[linewidth=1pt,arrowsize=3pt 2]{|-}(7.5,2.5)(7.5,3.8)
    \psline[linewidth=1pt,arrowsize=3pt 2]{-|}(7.5,4.2)(7.5,5.5)
    \put(7.35,3.9){$H$}
    \put(2.6,0.6){$x=0$}
    \put(5.6,0.6){$x=L$}
    \put(1,1.5){$\mbox{\Large $I$}$}
    \put(4.3,1.5){$\mbox{\Large $II$}$}
    \put(7.7,1.5){$\mbox{\Large $III$}$}
\end{picture}
\caption{Geometry of the 2-chain $\Omega$ used to calculate the time-averaged integrated pressure on the magnetoelectric slab.}
\end{center}
\label{slab}
\end{figure}
The above fields yield a net pressure on $II$ that fluctuates with
time, with a non-zero average.
If $A(\rr,t)$ is a scalar field, its average over any time interval $T$ is
\begin{eqnarray*}
    <A>(\rr) \equiv \frac{1}{T}\int_{0}^{T} A(\rr,t) dt.
\end{eqnarray*}
Hence, if $B(\rr,t)$ is another scalar field,
\begin{eqnarray*}
    <AB>(\rr) \equiv \frac{1}{T}\int_{0}^{T} A(\rr,t)B(\rr,t) dt.
\end{eqnarray*}
Furthermore, if
\begin{eqnarray*}
    \mathbf{A} &=& \Re\(\df {\cal A}(\rr)\exp(-i\omega t)\) \in \sect \Lambda^{p}M \\
    \mathbf{B} &=& \Re\(\df {\cal B}(\rr)\exp(-i\omega t)\) \in \sect \Lambda^{q}M,
\end{eqnarray*}
where ${\cal A,B}$ are complex, then
\begin{eqnarray*}
    \mathbf{A} \w \mathbf{B} &=& \frac{1}{2}\Re\(\df {\cal A}\w{\cal B}\exp(-2i\omega t)\) + \frac{1}{2}\Re({\cal A}\w\conj{\cal B}),
\end{eqnarray*}
so
\begin{eqnarray}
    \label{TA2F} <\mathbf{A} \w \mathbf{B}>(\rr) &=& \frac{1}{2}\Re({\cal A}\w\conj{\cal B}),
\end{eqnarray}
if we take $T=\frac{2\pi}{\omega}$. Thus, (\ref{SymMinkStress})
gives
\begin{eqnarray*}
    <\sigma_{K}^{U}>(\rr) &=& \frac{1}{4}\Re\( \df \Fee{}{y}(K)\# \conj{\Fdd{}{y}} + \Fdd{}{y}(K) \# \conj{\Fee{}{y}} + \Fhh{}{y}(K)\# \conj{\Fbb{}{y}} + \Fbb{}{y}(K)\# \conj{\Fhh{}{y}} \)   \\
    \nonumber  & & - \frac{1}{4}\#\Re\(\Fee{}{y} \w \# \conj{\Fdd{}{y}} +  \Fbb{}{y} \w \#\conj{\Fhh{}{y}} \)\#\wt{K}+ \frac{1}{4}\wt{U}(K)\Re\(\frac{1}{\cc}\Fee{}{y} \w \conj{\Fhh{}{y}} + \cc\Fdd{}{y} \w \conj{\Fbb{}{y}}\).
\end{eqnarray*}
Furthermore, with $U=\frac{1}{\cc}\partial_{t}$ and
$K=\partial_{x}$, the $x$-component of the time-averaged
Maxwell-Cauchy stress 2-form is
\begin{eqnarray*}
    <\sigma_{\partial_{x}}>(\rr) &=& \frac{1}{4}\Re\( \df \Fee{}{y}(\partial_{x})\# \conj{\Fdd{}{y}} + \Fdd{}{y}(\partial_{x}) \# \conj{\Fee{}{y}} + \Fhh{}{y}(\partial_{x})\# \conj{\Fbb{}{y}} + \Fbb{}{y}(\partial_{x})\# \conj{\Fhh{}{y}} \)  \\
    \nonumber  & & - \frac{1}{4}\#\Re\(\Fee{}{y} \w \# \conj{\Fdd{}{y}} +  \Fbb{}{y} \w \#\conj{\Fhh{}{y}} \)\#dx,
\end{eqnarray*}
which reduces to
\begin{eqnarray}
    \label{TASigma} <\sigma_{\partial_{x}}>(\rr) &=& -\frac{1}{4}\#\Re\(\Fee{}{y} \w \# \conj{\Fdd{}{y}} +  \Fbb{}{y} \w \#\conj{\Fhh{}{y}} \)dy \w dz.
\end{eqnarray}
The time-averaged integrated force is given by\footnote{The minus
sign occurs due to the opposite orientation of the opposite faces
of $\Omega$.}
\begin{eqnarray*}
    <f_{\partial_{x}}^{NET}[\Omega]> &=& \int_{\Omega_{0}} <\sigma_{\partial_{x}}^{I}> - \int_{\Omega_{L}} <\sigma_{\partial_{x}}^{III}>.
\end{eqnarray*}
Denote the time-averaged stress forms due to the fields in regions
$I$ and $III$ by
\begin{eqnarray*}
   <\sigma_{\partial_{x}}^{I}> &=& \alpha^{I} \, dy \w dz, \qquad \alpha^{I}=-\frac{1}{4}\#\Re\(\Fee{I}{y} \w \# \conj{\Fdd{I}{y}} +  \Fbb{I}{y} \w \#\conj{\Fhh{I}{y}} \) \\
   <\sigma_{\partial_{x}}^{III}> &=& \alpha^{III} \, dy \w dz, \qquad \alpha^{III}=-\frac{1}{4}\#\Re\(\Fee{III}{y} \w \# \conj{\Fdd{III}{y}} +  \Fbb{III}{y} \w \#\conj{\Fhh{III}{y}} \)
\end{eqnarray*}
Thus, the time-averaged net force on the magnetoelectric medium
contained in the region bounded by $\Omega$ is
\begin{eqnarray*}
    <f^{NET}_{\partial_{x}}[\Omega]> &=&  \(\Omega^{*}_{0}\alpha^{I} - \Omega^{*}_{L}\alpha^{III}\)\int_{0}^{W}\int_{0}^{H}dy dz \= \(\Omega^{*}_{0}\alpha^{I} - \Omega^{*}_{L}\alpha^{III}\)A,
\end{eqnarray*}
where $A=WH$ and the time-average integrated pressure
$<{p}_{x}[\Omega]>\,\equiv\frac{<f^{NET}_{\partial_{x}}[\Omega]>}{A}$.
Calculating the pull-backs of
\begin{eqnarray*}
    \alpha^{I} &=& -\frac{|{\cal E}|^{2}}{4}\[ \ep{0}\( 1 + 2\Re(\FE{I}{L}{y}\exp(i [\Fk{I}{R}{y} - \Fk{I}{L}{y}]x )) + |\FE{I}{L}{y}|^{2} \) \frac{}{} \right. \\ & & \left. \hspace{1.9cm} + \frac{1}{\mu_{0}\omega^{2}}\((\Fk{I}{R}{y})^2 + 2\Fk{I}{R}{y}\Fk{I}{L}{y}\,\Re(\FE{I}{L}{y}\exp(i [\Fk{I}{R}{y} - \Fk{I}{L}{y}]x )) + (\Fk{I}{L}{y})^2 |\FE{I}{L}{y}|^{2} \) \] \\
    && \\
    \alpha^{III} &=& -\frac{|{\cal E}\FE{III}{R}{y}|^{2}}{4}\( \ep{0} + \frac{(\Fk{III}{R}{y})^{2}}{\mu_{0}\omega^{2}}  \)
\end{eqnarray*}
yields
\begin{eqnarray*}
    \Omega^{*}_{0}\alpha^{I} &=& -\frac{\ep{0}|{\cal E}|^{2}}{2}\(  1  + |\FE{I}{L}{y}|^{2} \) , \qquad \Omega^{*}_{L}\alpha^{III} \= -\frac{\ep{0}|{\cal E}\FE{III}{R}{y}|^{2}}{2},
\end{eqnarray*}
since
$\Fk{I}{R}{y}=-\Fk{I}{L}{y}=\Fk{III}{R}{y}=\frac{\omega}{\cc}$.
Since the time-averaged body force $<\Lie_{U}\rho^{U}_{K}>=0$ for $\rho^{U}_{K}$ given by (\ref{AbrD}), (\ref{MinkD}) and (\ref{SymMinkD}), it follows that the average pressure on the magnetoelectric slab is given
in terms of the solution (\ref{fieldsols}) by
\begin{eqnarray}
    \label{Pressure} <{p}_{x}[\Omega]> &=& \frac{\ep{0}|{\cal E}|^{2}}{2}\(  \df |\FE{III}{R}{y}|^{2} - |\FE{I}{L}{y}|^{2} - 1  \, \).
\end{eqnarray}\\

\section{Conclusion}
\label{discussion}
The magnitude and sign of $<{p}_{x}[\Omega]>$ depends on
$\ep{}\equiv\ep{r}\ep{0}, \mu\equiv\mu_r\mu_0, \beta_1$ and
$\beta_2$, where $\cc=\frac{1}{\sqrt{\ep{0}\mu_0}}$. As noted
above, the wave numbers $\Fk{II}{L}{y} , \Fk{II}{R}{y}$ that
follow from the dispersion relation determine the nature of the
propagating wave in region $II$. For the case under discussion
here, where the parameters $\ep{r}, \mu_r, \beta_1,\beta_2$ are
constant, it is of interest to write the dispersion relations in
terms of the dimensionless ratio of the wave speeds $w\equiv
\frac{v}{v_0}$, where $v=\frac{\omega}{k}$,
$v_0=\frac{1}{\sqrt{\ep{}\mu}}$ and the dimensionless parameters
$b_1\equiv -\beta_1/\sqrt{\ep{}\mu},\,b_2\equiv
\beta_2/\sqrt{\ep{}\mu}$:
\begin{eqnarray*}
    w^2 + 2 b_1 w - 1 &=& 0 \\
    w^2 + 2 b_2 w - 1 &=& 0.
\end{eqnarray*}
Then the sub-luminal condition $| \frac{v}{\cc} | < 1$ implies $|
w | < \sqrt{\ep{r}\mu_r}$. The relation between $w$ and either
$b_1$ and $b_2$ can then be seen from the relation of the two
branches of the loci where the expression $w^2 + 2 b w-1$ vanishes
in the $w$-$b$ plane. For $\omega>0$, values of $w$ in the upper
(lower) half plane correspond to left (right) moving waves.
Furthermore, propagating sub-luminal monochromatic waves will only
occur in $II$ for real $b$, yielding real values of $w$ in the
range $-\sqrt{\ep{r}\mu_r} < w < \sqrt{\ep{r}\mu_r}$. It is clear
from these considerations that the relative sign between $\beta_1$
and $\beta_2$ can have a significant effect on the behavior of the
propagating modes in the region $II$ and hence on the nature of
the force on the magnetoelectric slab.

The authors feel that the approach adopted in this paper for the
calculation of static, time averaged and instantaneous forces,  offers a conceptually unambiguous
method of considerable generality. Once one decides on the drive
form appropriate for any subsystem in interaction with external
fields, it has immediate application to moving media (in arbitrary
relativistic or non-relativistic motion) and can be extended to
matter with material losses. Work is in progress to extend the
methodology to inhomogeneous media with more general constitutive
properties and this will be reported elsewhere.\\

\acknowledgments The authors thank R. Seviour for useful
discussions and  are grateful to EPSRC and the Cockcroft Institute
for supporting this research.\\

\appendix
\section*{Electromagnetic Fields in the Three Regions}
\label{AppMEslab}
For a $y$-polarised harmonic electromagnetic wave with angular frequency $\omega > 0$, incident normally from
the left on a fixed magnetoelectric slab, the electric field solutions in the three regions are
given by (\ref{FeeI})$-$(\ref{FeeIII}). For completeness, the
remaining fields in these three regions are given here. The
magnetic induction fields follow from (\ref{F1}):
\begin{eqnarray*}
    \Fbb{I}{y} &=& \frac{\EU \Fk{I}{R}{y}}{\omega}\exp(i \Fk{I}{R}{y}  x -i\omega\, t ) \, dz + \frac{\EU\Fk{I}{L}{y}\FE{I}{L}{y}}{\omega} \,\exp(i\Fk{I}{L}{y} x-i\omega\, t)\,dz  \\
    \Fbb{II}{y} &=& \frac{\EU\Fk{II}{R}{y}\FE{II}{R}{y}}{\omega} \exp(i\Fk{II}{R}{y} x -i\omega\, t)\,dz + \frac{\EU\Fk{II}{L}{y}\FE{II}{L}{y}}{\omega}\exp({i \Fk{II}{L}{y} x-i\omega\, t  })\, dz \\
    \Fbb{III}{y} &=& \frac{\EU\Fk{III}{R}{y}\FE{III}{R}{y}}{\omega} \exp(i \Fk{III}{R}{y} x-i\omega\, t)\, dz.
\end{eqnarray*}
The electric displacement 1-forms in regions $I$ and $III$ are
given by the vacuum constitutive relation
$\Fdd{}{y}=\ep{0}\Fee{}{y}$, whereas the electric displacement
1-form in region $II$ is given by the constitutive relation
(\ref{FCR1}), with the spatial tensors $\Xde$ and $\Xdb$ given by
(\ref{XdeMS}) and (\ref{XdbMS}) respectively:
\begin{eqnarray*}
    \Fdd{I}{y} &=& \EU\ep{0}\exp(i \Fk{I}{R}{y}  x -i\omega\, t ) \, dy + \EU\ep{0}\FE{I}{L}{y} \,\exp(i\Fk{I}{L}{y} x-i\omega\, t)\,dy \\
    \Fdd{II}{y} &=&\EU \(\ep{} + \frac{\beta_{2}\Fk{II}{R}{y}}{\omega} \)\FE{II}{R}{y} \exp(i\Fk{II}{R}{y} x -i\omega\, t)\,dy + \EU\(\ep{} + \frac{\beta_{2}\Fk{II}{L}{y}}{\omega} \)\FE{II}{L}{y} \exp({i \Fk{II}{L}{y} x-i\omega\, t  })\, dy \\
    \Fdd{III}{y} &=& \EU\ep{0}\FE{III}{R}{y} \exp(i \Fk{III}{R}{y} x-i\omega\, t)\, dy.
\end{eqnarray*}
Similarly, the magnetic 1-forms in the regions $I$ and $III$ are
given by the vacuum constitutive relation
$\Fhh{}{y}=\mu_{0}^{-1}\Fbb{}{y}$, whereas in region $II$, the
magnetoelectric constitutive relation (\ref{FCR2}), with the
spatial tensors $\Xhb$ and $\Xhe$ given by (\ref{XhbMS}) and
(\ref{XheMS}) respectively yield
\begin{eqnarray*}
    \Fhh{I}{y} &=& \frac{\EU\Fk{I}{R}{y}}{\mu_{0}\omega}\exp(i \Fk{I}{R}{y}  x -i\omega\, t ) \, dz + \frac{\EU\Fk{I}{L}{y}\FE{I}{L}{y}}{\mu_{0}\omega} \,\exp(i\Fk{I}{L}{y} x-i\omega\, t)\,dz  \\
    \Fhh{II}{y} &=& \EU\(\frac{\Fk{II}{R}{y}}{\mu\omega} - \beta_{2} \)\FE{II}{R}{y}\exp(i\Fk{II}{R}{y} x -i\omega\, t)\,dz + \EU\(\frac{\Fk{II}{L}{y}}{\mu\omega} - \beta_{2} \)\FE{II}{L}{y}\exp({i \Fk{II}{L}{y} x-i\omega\, t  })\, dz \\
    \Fhh{III}{y} &=& \frac{\EU\Fk{III}{R}{y}\FE{III}{R}{y}}{\mu_{0}\omega} \exp(i \Fk{III}{R}{y} x-i\omega\, t)\, dz.
\end{eqnarray*}




\begin{thebibliography}{0}
    \bibitem{Maugin} \BY{G.A. Maugin}
    \TITLE{On the Covariant Equations of the Relativistic Electrodynamics of Continua. III. Elastic Solids},
                  \IN{J. Math. Phys.}{19}{1978}{1212-1219}

    \bibitem{Landau} \BY{L. D. Landau, E. M. Lifschitz \atque L. P. Pitaevskii}
    \TITLE{Electrodynamics of Continuous Media - Volume 8 in Course of Theoretical Physics},
                  (Butterworth-Heinemann) 1984

    \bibitem{Barnett} \BY{S. M. Barnett \atque R. Loudon}
    \TITLE{On the Electromagnetic Force on a Dielectric Medium},
                  \IN{J. Phys. B: At. Mol. Opt. Phys.}{39}{2006}{671-684}

    \bibitem{Mog} \BY{E. R. Mognaschi \atque A. Savin}
   \TITLE{The Action of a Non-Uniform Electric Field Upon a Lossy Dielectric Systems $-$ Ponderomotive Force on a Dielectric
    Sphere in the Field of a Point Charge},
                  \IN{J Phys. D: Appl. Phys.}{16}{1983}{1533-1541}

    \bibitem{Giner} \BY{V. Giner, M. Sancho \atque G. Martinez}
    \TITLE{Electromagnetic Forces on Dissipative Dielectric Media},
                  \IN{Am. J. Phys.}{63}{1995}{749-753}

    \bibitem{Mans} \BY{M. Mansuripur}
    \TITLE{Electromagnetic Force and Torque in Ponderable Media},
                  \IN{Optics Express}{16}{2008}{14821-14835}


    \bibitem{RWT_JMP} \BY{R. W. Tucker}
    \TITLE{Differential Form Valued Forms and Distributional Electromagnetic Sources},
                  \IN{ arXiv:0812.1959v1,  J. Math. Phys. To Appear}{}{2009}{}


    \bibitem{Dietz} \BY{E. R. Dietz}
    \TITLE{Force on a Dielectric Slab: Fringing Field Approach},
                  \IN{Am. J. Phys.}{72}{2004}{1499-1500}

    \bibitem{Hinds} \BY{E. A. Hinds \atque S. M. Barnett}
    \TITLE{Momentum Exchange Between Light and a Single Atom: Abraham or Minkowski?},
                  \IN{arXiv:0811.2771}{}{2008}{1-4}

    \bibitem{Sim} \BY{C.R. Simovski}
    \TITLE{Bloch Material Parameters of Magneto-Dielectric Metamaterials and the Concept of Bloch Lattices},
                  \IN{Metamaterials}{1}{2007}{62-80}

    \bibitem{RWT} \BY{I. M. Benn \atque R. W. Tucker}
    \TITLE{An Introduction to Spinors and Geometry with Applications in Physics},
                  (Adam Hilger: IoP Publishing) 1988

    \bibitem{Shilov} \BY{G. E. Shilov}
    \TITLE{Generalized Functions and Partial Differential Equations},
                  (Gordon and Breach) 1968

    \bibitem{Benn} \BY{I. M. Benn}
    \TITLE{Conservation Laws in Arbitrary Space-times},
                  \IN{Ann. Inst. H. Poincar\'{e}}{37}{1982}{67-91}

    \bibitem{ODell} \BY{T. H. O'Dell}
    \TITLE{The Electrodynamics of Magneto-electric Media,
                  (North-Holland) 1970}

    \bibitem{Pfeifer} \BY{R. N. C. Pfeifer, T. A. Nieminen, N. R. Heckenberg \atque H. Rubinsztein-Dunlop}
    \TITLE{Colloquium: Momentum of an Electromagnetic Wave in Dielectric Media},
                  \IN{Rev. Mod. Phys.}{79}{2007}{1197-1216}

    \bibitem{Maugin1} \BY{G.A. Maugin}
    \TITLE{Further comments on the equivalence of Abraham's Minkowski's and Others' electrodynamics},
                  \IN{Can. J. Phys.}{58}{1980}{1163-1170}

    \bibitem{DGT} \BY{T. Dereli, J. Gratus \atque R. W. Tucker}
    \TITLE{The Covariant Description of Electromagnetically Polarizable Media},
                  \IN{Phys. Lett. A}{361}{2006}{190-193}

    \bibitem{DGT2} \BY{T. Dereli, J. Gratus \atque R.W. Tucker}
    \TITLE{New Perspectives on the Relevance of Gravitation for the Covariant Description
    of Electromagnetically Polarizable Media},
                  \IN{J. Phys A: Math. Theor.}{10}{2007}{5695-5715}

    \bibitem{HO} \BY{Y. N. Obukhov and F. W. Hehl}
    \TITLE{Forces and Momenta Caused by Electromagnetic Waves in Magnetoelectric Media},
                  \IN{Phys. Letts. A}{372}{2008}{3946-3952}

    \bibitem{Jackson} \BY{J. D. Jackson}
    \TITLE{Classical Electrodynamics},
                  (Wiley, 3rd Edition) 1998



\end{thebibliography}
\end{document}